\documentclass{ws-ijmpe}

\usepackage{graphicx}
\usepackage{epsfig}
\usepackage{amsfonts}
\usepackage{amsmath,amssymb}
\usepackage{subfig}

\begin{document}

\markboth{M. Zalewski, P. Olbratowski, and W. Satu{\l}a}
{The nuclear energy density
functionals with modified...
}

%
\catchline{}{}{}{}{}
%

\title{The nuclear energy density
functionals with modified radial dependence of the
isoscalar effective mass}

\author{M. ZALEWSKI\footnote{Maciej.Zalewski@fuw.edu.pl},~~P.
OLBRATOWSKI,~~W. SATU{\L}A\\}

\address{Institute of Theoretical Physics, University of Warsaw,
ul. Ho\.za 69, 00-681 Warsaw, Poland}

\maketitle
\begin{history}
\received{(received date )}
\revised{(revised date )}
\end{history}

\begin{abstract}
Calculations for infinite nuclear matter with realistic nucleon-nucleon interactions suggest that the isoscalar effective mass (IEM) of a nucleon at the saturation density equals $m^*/m\sim 0.8\pm 0.1$, at variance with empirical data on the nuclear level density in finite nuclei which are consistent with  $m^*/m\approx 1$.
This contradicting results might be reconciled by enriching the radial
dependence of IEM. In this work four new terms are introduced into the
Skyrme-force inspired local energy-density functional: $\tau(\nabla\rho)^2$,
$\tau\frac{d\rho}{dr}$, $\tau^2$ and $\tau\Delta\rho$. The aim is to
investigate how they influence the radial dependence of IEM and, in turn,
the single-particle spectra. \end{abstract}

\section{Introduction}

The nuclear energy-density functional (EDF) is considered nowadays as one
of the most promising theoretical tools to describe static properties of
the atomic nuclei. Tremendous effort is undertaken recently to develop
high-precision spectroscopic quality
functionals. A standard way to construct the nuclear EDF is to start with
either the finite-range Gogny \cite{[Gog75aw]} or the zero-range Skyrme \cite{[Sky56xw]} effective
interaction and average it with the density matrix within the Hartree-Fock
(HF) method. Such functionals may be further enriched by adding new terms. For
instance, Carlsson {\it et al.} \cite{[Car08]} considered a systematic
generalization of the Skyrme functional by introducing terms up to sixth order
in derivatives of the density matrix.

Calculations for infinite nuclear matter suggest that the isoscalar effective mass (IEM) at the saturation density should be
of order of $m^*/m \sim 0.8 \pm 0.1$~\cite{[Bru58],[Jeu76w],[Fri81],[Wir88],[Zuo99]}.
On the other hand, description of the single-particle (s.p.) level density in
finite nuclei requires the IEM close to unity. This contradicting conditions
might be fulfilled together by assuming the IEM smaller than one inside the
nucleus and peaked at the surface. Such a concept was first explored by Ma and
Wambach \cite{[Ma83]} in a non-self-consistent model and by Farine {\it et
al.}~\cite{[Far01]} within a self-consistent framework. Since it is
impossible to obtain a surface-peaked IEM within the standard form of
Skyrme functional, in the present paper we examine four simple additional
terms that can modify the radial dependence of IEM.

The new terms are introduced in Sec. \ref{terms}. Sec. \ref{dziubek} reports on the resulting
radial dependence of the IEM. In Sec. \ref{spe}, influence of the new terms on
the s.p. spectra is presented. The paper is concluded in Sect.~\ref{summary}.
In this exploratory work, we limit our calculations to three spherically
symmetric isoscalar nuclei, $^{40}$Ca, $^{56}$Ni, and $^{100}$Sn.

\section{Extensions to the Skyrme energy-density functional}\label{terms}

The Skyrme energy density, ${\mathcal H}({\mathbf r})$, consists of the kinetic and interaction parts,
\begin{equation}\label{eq108}
   {\mathcal H}({\mathbf r}) = \frac{\hbar^2}{2m}\tau_0
               +  \displaystyle\sum_{t=0,1} {\mathcal H}_t({\mathbf r}) ,
   \end{equation}
where
\begin{equation}
\label{hte}
\mathcal{H}_t=C^{\rho}_t[\rho_0]\rho^2_t+C^{\Delta\rho}_t\rho_t\Delta\rho_t+C^{\tau}_t\rho_t\tau_t+C^J_t{\mathbb J}^2_t+C^{\nabla J}_t\rho_t{\mathbf\nabla}\cdot{\mathbf J}_t.
\end{equation}
Index $t=0,1$ denotes the isospin, and $C_t$ are the coupling constants, of which one,
$C^{\rho}_t[\rho_0]=C^{\rho\prime}_t+C^{\rho\prime\prime}_t
{\rho_0}^{\alpha}$, depends on the isoscalar density. The potential energy
terms are bilinear forms of the time-even densities, $\rho$, $\tau$, ${\mathbb
J}$, and their derivatives. The density ${\mathbf J}_{t}$ denotes the vector
part of the spin-current tensor, ${\mathbf
J}_{t,\lambda}=\sum_{\mu\nu}\epsilon_{\lambda\mu\nu}{\mathbb J}_{t,\mu\nu}$.
Readers interested in details are referred, e.g., to Ref.~\cite{[Ben03]}.

The central field, $\mathcal{U}({\mathbf r})$, is defined as a variation of the energy density with respect to the particle density,
\begin{equation}
\mathcal{U}_t({\mathbf r})=\frac{\delta [ \mathcal{H}({\mathbf r}) ]}{\delta \rho_t({\mathbf r})}
= 2C^{\rho}_t[\rho_0] \rho_t
+ 2C^{\Delta \rho}_t \Delta\rho_t
+ C^{\tau}_t\tau_t
+ C^{\nabla J}_t {\mathbf \nabla}\cdot{\mathbf  J}_t
+ \delta_{t0}\, \sum_{t^\prime =0,1} \frac{\partial
C^{\rho}_{t^\prime}}{\partial\rho_0} \rho_{t^\prime}^2.
\end{equation}
The effective
mass is proportional to the inverse of the mass field, which is a variation of
the energy density with respect to the kinetic density,
\begin{equation}\label{m-sky} \frac{\hbar^2}{2m_t^*({\mathbf
r})}=\mathcal{M}_t({\mathbf r})=\frac{\delta [ \mathcal{H}({\mathbf
r})]}{\delta \tau_t({\mathbf r})}=\frac{\hbar^2}{2m}\delta_{t0} + C^{\tau}_t
\rho_t ({\mathbf r}).
\end{equation}
The IEM is defined as $m_0^*/m$. It is easily seen from Eq.~(\ref{m-sky}) that
in the standard Skyrme functional, the IEM is a monotonic function of the
isoscalar particle density, $\rho_0$, and
therefore cannot be peaked at the surface. The term proportional to $\rho_0$
in Eq.~(\ref{m-sky}) can only decrease or increase the IEM in the interior of
the nucleus.

Thus, in order to enrich the radial dependence of the IEM, one has to add to the functional new terms depending on the isoscalar kinetic density, $\tau_0$.
We consider here four such terms,
\begin{eqnarray}
{\mathcal H}_0^{(A)}({\mathbf r}) & = & C_0^{\tau (\mathbf{\nabla}\rho)^2}
\tau_0 ({\mathbf r}) \left( \mathbf{\nabla}\rho_0 ({\mathbf r}) \right)^2 ,\label{tgr} \\
{\mathcal H}_0^{(B)}({\mathbf r}) & = & C_0^{\tau \frac{d\rho}{dr}}
\tau_0 ({\mathbf r}) \frac{d\rho_0 ({\mathbf r}) }{dr}, \label{tdr} \\
{\mathcal H}_0^{(C)}({\mathbf r}) & = & C_0^{\tau ^2}
(\tau_0 ({\mathbf r}))^2, \label{tsq} \\
{\mathcal H}_0^{(D)}({\mathbf r}) & = & C_0^{\tau \Delta\rho}
\tau_0 ({\mathbf r}) \Delta \rho_0({\mathbf r}) . \label{tlr}
\end{eqnarray}
These terms will be treated independently and dubbed
as variants A, B, C, and D of our model, respectively.
Their contributions to the mass field read, respectively,
\begin{eqnarray}
{\mathcal M}_0^{(A)}({\mathbf r}) & = & C_0^{\tau (\mathbf{\nabla}\rho)^2}
 \left( \mathbf{\nabla}\rho_0 ({\mathbf r}) \right)^2 ,\label{m-tgr} \\
{\mathcal M}_0^{(B)}({\mathbf r}) & = & C_0^{\tau \frac{d\rho}{dr}}
 \frac{d\rho_0 ({\mathbf r}) }{dr}, \label{m-tdr} \\
{\mathcal M}_0^{(C)}({\mathbf r}) & = & 2C_0^{\tau ^2}
\tau_0 ({\mathbf r}), \label{m-tsq} \\
{\mathcal M}_0^{(D)}({\mathbf r}) & = & C_0^{\tau \Delta\rho}
 \Delta \rho_0({\mathbf r}) \label{m-tlr}.
\end{eqnarray}
All of these terms but ${\mathcal H}_0^{(C)}$ contribute to the central field as well,
\begin{eqnarray}
{\mathcal U}_0^{(A)}({\mathbf r}) & = & -2C_0^{\tau (\mathbf{\nabla}\rho)^2}
 \left( \mathbf{\nabla}\tau_0  \cdot \mathbf{\nabla}\rho_0 +
               \tau_0  \Delta \rho_0  \right) ,\label{u-tgr} \\
{\mathcal U}_0^{(B)}({\mathbf r}) & = & C_0^{\frac{d\rho }{dr}}\frac{d\tau_0 }{dr}, \label{u-tdr} \\
{\mathcal U}_0^{(D)}({\mathbf r}) & = & C_0^{\tau \Delta\rho}
 \Delta \tau_0({\mathbf r}) . \label{u-tlr}
\end{eqnarray}

The terms A, C, and D are scalars.
Hence, their form is natural for all shapes of the nucleus.
The term B is valid only in case of spherical symmetry.
It is inspired by the Ma and Wambach parameterization of IEM.

The terms A and B depend on the first derivative of the particle density. This derivative is peaked at the surface, so these terms are expected to yield the desired profile of the IEM. Since the gradient square is positive and the radial derivative itself is negative, an upward-pointing peak will require negative and positive values of the coupling constants in cases A and B, respectively.

The term D depends on the second derivative. Hence, it should
produce two peaks of opposite signs at the borders of the surface, where the
curvature of the density profile is largest.

\section{Radial profiles of the isoscalar effective mass}\label{dziubek}

Correct description of the s.p.-level density near the Fermi surface requires the mean value of the IEM
to be close to unity.
In order to fulfill this condition we impose the following constraint on the radial IEM profile,
\begin{equation}\label{m-unit}
 \int d^3 {\mathbf r}\, \frac{\rho_0({\mathbf r})}{A}\, \frac{m^*({\mathbf r})}{m} = 1,
\end{equation}
where $A$ is the mass number of the considered nucleus.

We examine each of the four new terms, A-D, separately, that is, by switching the remaining three off.
In each case, we vary the concerned new coupling constant by hand, and for each value thereof, we readjust
the coupling constant $C^{\tau}_0$ to fulfill the condition (\ref{m-unit}).
This means that the excess of the IEM produced by the peak is compensated by decreasing the IEM inside the nucleus.

The remaining coupling constants of the functional (\ref{hte}) are kept intact at values of the SkXc Skyrme
parameterization~\cite{[Bro98]}. This specific parameterization was chosen
because of the condition (\ref{m-unit}). Indeed,
in order to fulfill Eq.~(\ref{m-unit}) it is reasonable to choose a
parameterization with the IEM close to unity, which is a case for the SkXc. In
addition, this force was fitted with particular attention payed to the s.p.
spectra.

\begin{figure}[t]
\begin{center}
\includegraphics[width=0.4\textwidth, clip]{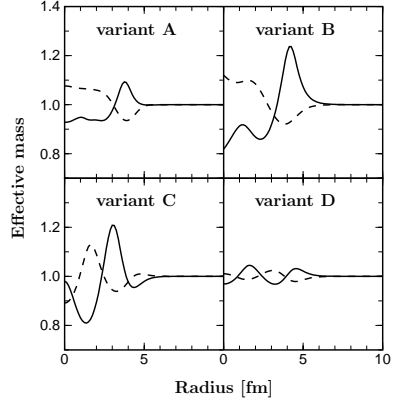}
\end{center}
\caption{Radial dependence of the IEM,
$m_0^*(r)/m$, in $^{40}$Ca for the four variants of the calculations.
The solid and dashed lines show the IEM profiles for the limiting achievable
values of the corresponding coupling constants. See text for details.}
\label{shapes} \end{figure}

It turned out that the possible values of the new coupling constants are limited.
Extending the new coupling constants
beyond certain limits leads to strong oscillations of the density
what causes that the total energy diverges.  The radial profiles of the IEM
obtained in $^{40}$Ca for the limiting (both negative and positive) values
of the new coupling constants are shown in Fig.~\ref{shapes}.
The limits corresponding to downward-pointing
peaks in the IEM are, most likely, of limited physical interest.

It is clearly seen that in the variants A and B of our model the desired
surface-peaked IEM profile is obtained. In the variant C a well pronounced
peak appears as well, but it is shifted toward the center of the nucleus.
The variant D yields only small fluctuations of the IEM value, and it does not
seem to be of much relevance.

\section{Influence of the new terms on single-particle energies}\label{spe}

\begin{figure}[t]
\begin{center}
\subfloat[]{\includegraphics[width=0.4\textwidth, clip]{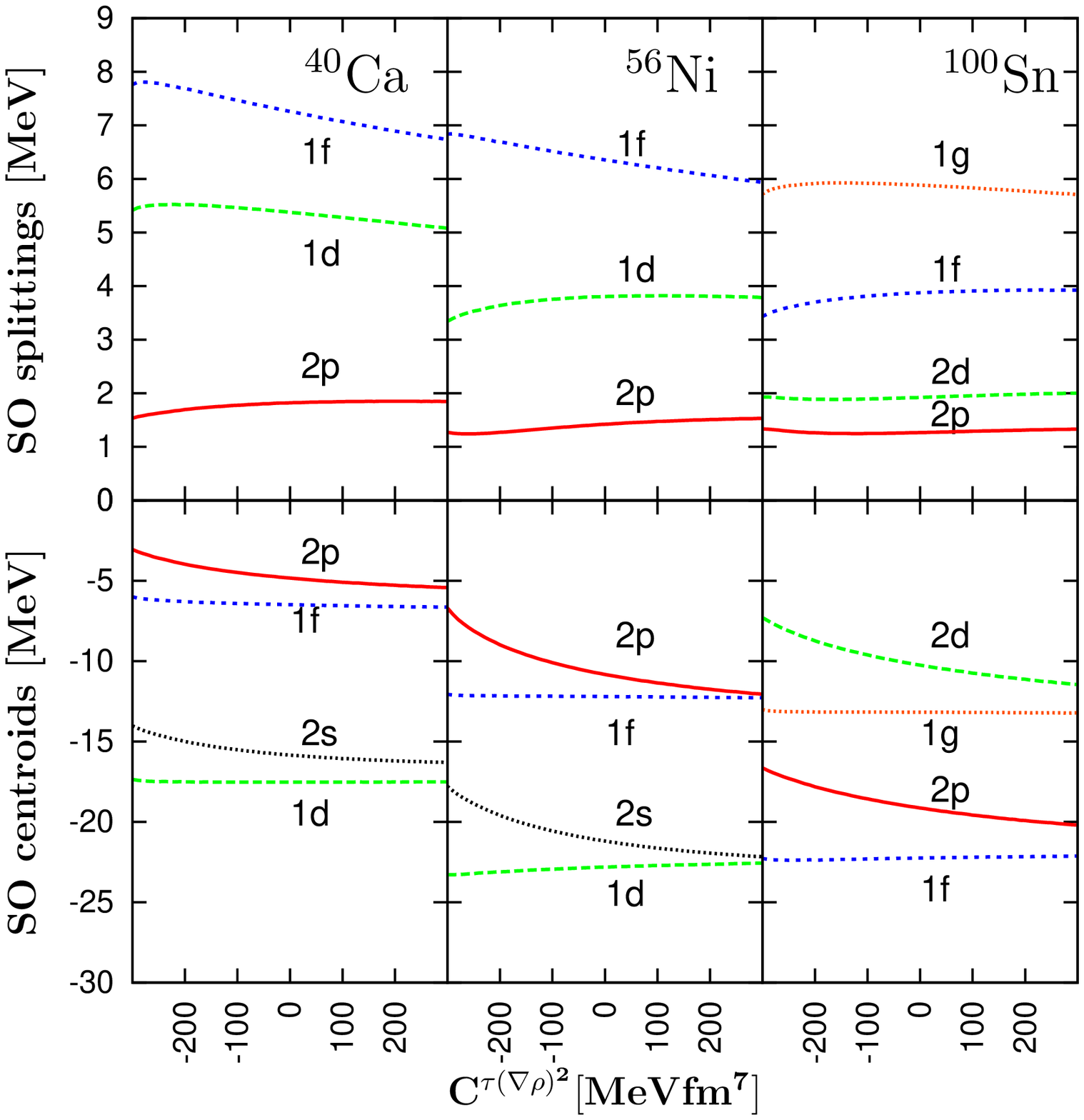}}\label{split_tdr2a}
\subfloat[]{\includegraphics[width=0.4\textwidth, clip]{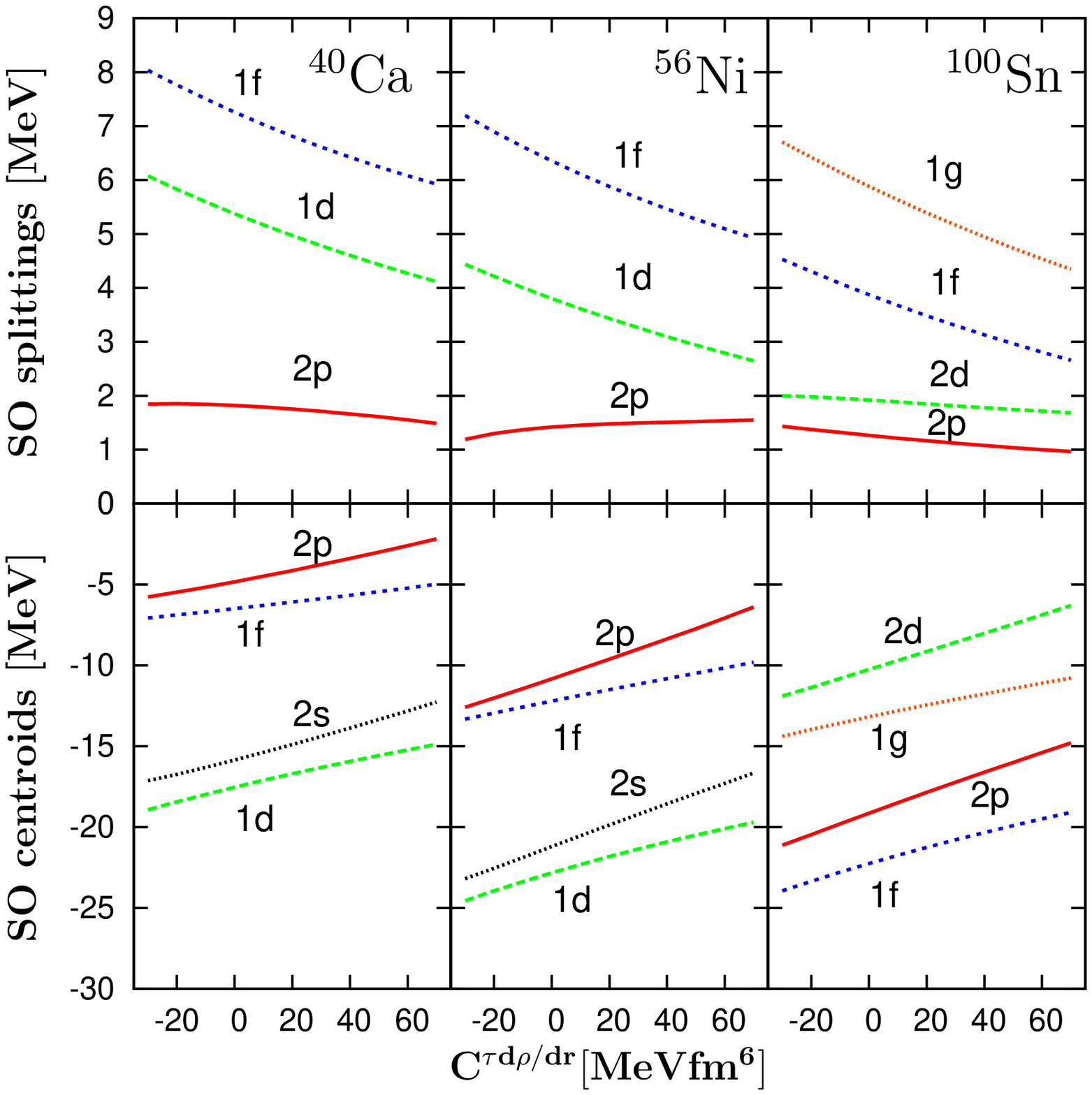}}\label{split_tdr2b}
\subfloat[]{\includegraphics[width=0.4\textwidth, clip]{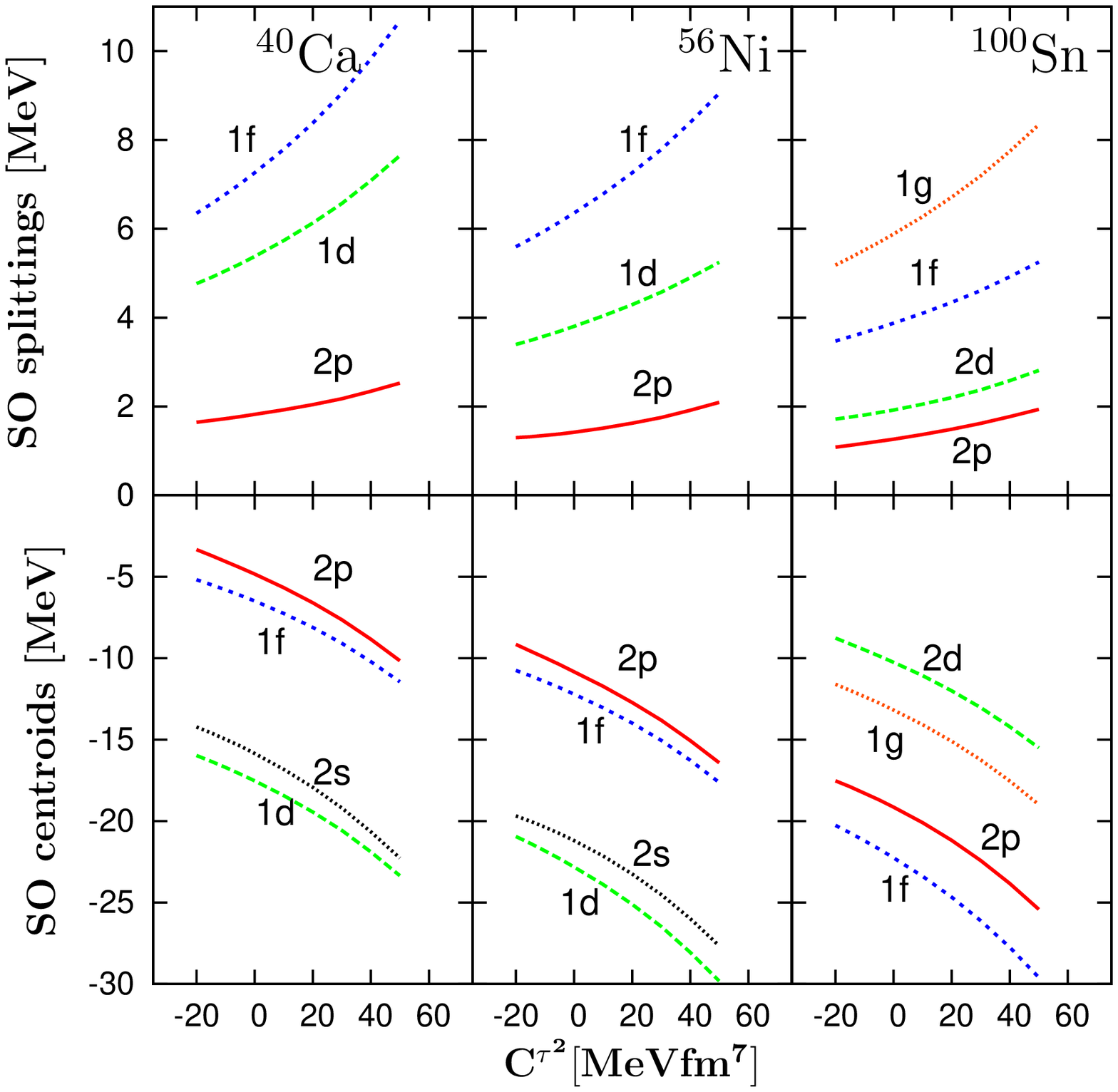}}\label{split_tdr2c}
\subfloat[]{\includegraphics[width=0.4\textwidth, clip]{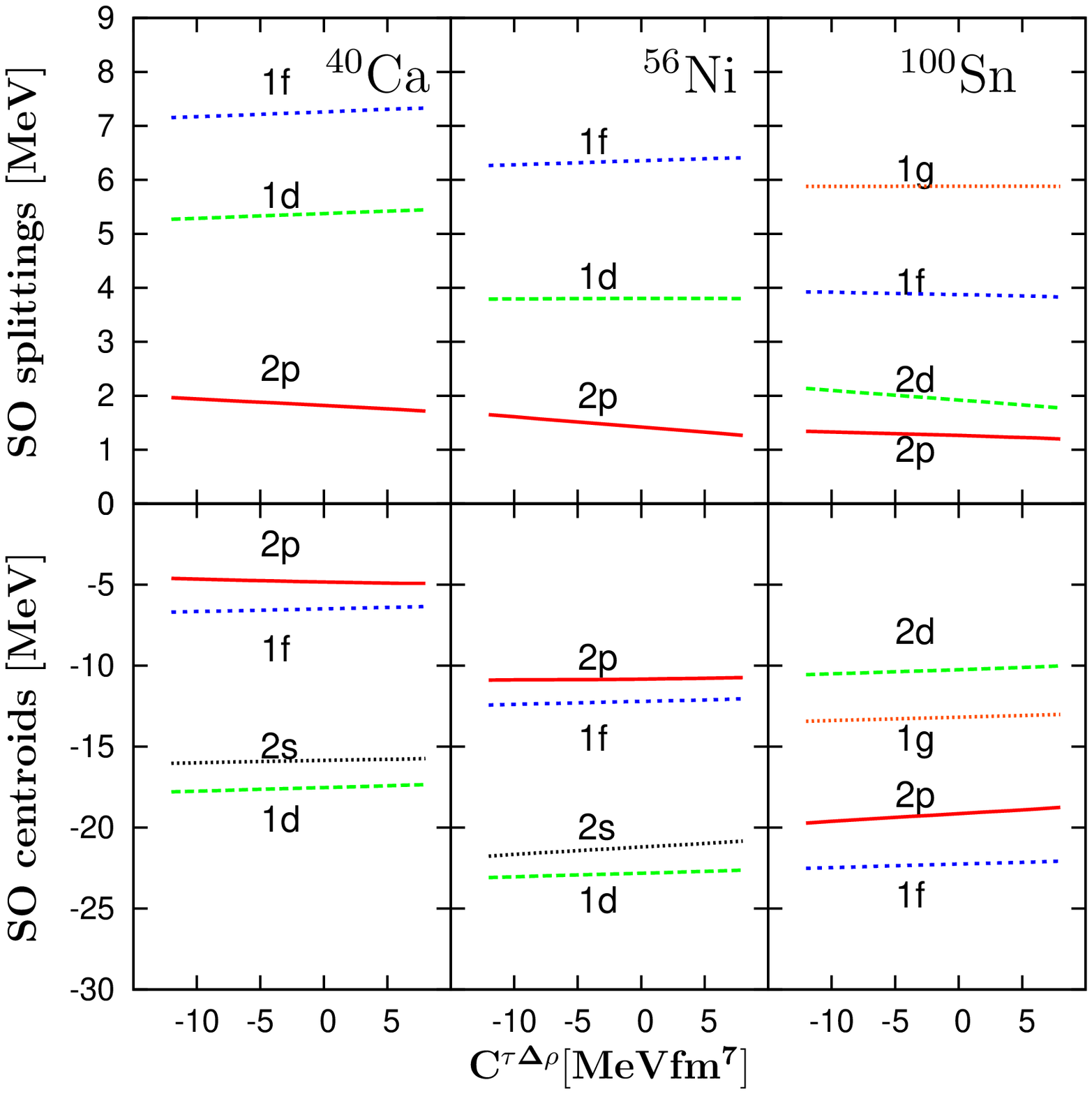}}\label{split_tdr2d}
\caption{Spin-orbit splittings (upper panels) and centroids (lower panels) as a function of
$C^{\tau\frac{d\rho}{dr}}$ (a) and $C^{\tau\frac{d\rho}{dr}}$ (b), $C^{\tau^2}$ (c) and  $C^{\tau\Delta\rho}$ (d)
in $^{40}$Ca (left), $^{56}$Ni (middle) and
$^{100}$Sn (right) nuclei.}
\label{spe-fig}
\end{center}
\end{figure}

In our study, we are mostly interested in the influence of the new terms on
the s.p. spectra, in particular on the spin-orbit (SO) coupling. For each pair
of the SO partners, with the same $n$, $l$, and $j=l\pm\frac{1}{2}$, we define
the SO {\it splitting} as the difference of their energies, and the SO {\it
centroid} as the arithmetic average thereof. Figure \ref{spe-fig} shows,
separately for each variant of our model, the SO splittings and centroids
as a function of new coupling constant for the entire accessible range of each
coupling constant.

The new terms influence both the SO splittings and centroids quite strongly in all
variants of our model except the variant D. In the latter case, the
modifications of the IEM's radial profile are apparently too weak to produce
any clear trend. In our calculations, the term C affects the s.p. levels most
strongly. However, it also shifts the binding energies of the considered
nuclei by more than 50\% when going from $C^{\tau^2}_0=0$ to the limiting
values. It is, therefore, clear that in order to remain within a physically
acceptable area, it would be necessary to refit all the coupling constants of
the functional at least to masses and radii of the three considered nuclei.
This is, however, beyond the scope of the present work. Such a problem
is of slightly lesser importance for the terms A and B, since their
inclusion does not change the masses by more than 10\%.

As expected, the terms A and B give a surface-peaked IEM profile for negative and
positive values of their coupling constants, respectively. Taking this into account,
one can see from Figs. \ref{spe-fig} (a) and (b) that the impact of the peak on the
SO splittings is opposite in these two cases, although the plots look similarly.
For the term A, the emergence of the peak leads to an enhancement of the SO
splitting of the $1f$ level in $^{40}$Ca. This trend is undesired since all
the existing Skyrme parameterizations overestimate this splitting already without
additional terms, see Ref.~\cite{[Zal08]} and articles quoted therein. On the other hand,
the onset of the peak caused by the term B quenches the $1f$ splitting in $^{40}$Ca,
driving it closer to the experimental value.

\section{Summary and conclusions}\label{summary}

We presented a way to extend the standard Skyrme EDF by introducing new terms:
$\tau(\nabla\rho)^2$, $\tau\frac{d\rho}{dr}$, $\tau^2$, and $\tau\Delta\rho$.
The rationale behind is to
modify the radial profile of the IEM toward a surface-peaked geometry.
In this way one can effectively include, within the nuclear EDF theory,
coupling to surface vibrations and restitute correct level density
at the Fermi surface~\cite{[Ber80w]}. Our study shows, in particular, that the
term $\tau\frac{d\rho}{dr}$ gives the desired surface-peaked IEM and that
it seems to be most promising as far as reproducing the experimental s.p. spectra is
concerned. However, this term is not a scalar, and must be
therefore generalized in order to be applicable to non-spherical shapes. A
more detailed analysis of the variants A and B of our model is underway.

\section{Acknowledgments}\label{acknowledgements}

We would like to thank Janusz Skalski for inspiring comments.
This work was supported in part by the Polish Ministry of Science under
Contracts No.~N~N202~239137 and~N~N202~328234.



\end{document}